# Escaping from air pollution: The psychological process of domestic migration intention among urban people

Quan-Hoang Vuong [1], Tam-Tri Le [1], Quang-Loc Nguyen [2], Quang-Trung Nguyen [3], and Minh-Hoang Nguyen [1,*]

[1] Centre for Interdisciplinary Social Research, Phenikaa University, Yen Nghia Ward, Ha Dong District, Hanoi 100803, Vietnam

[2] College of Asia Pacific Studies, Ritsumeikan Asia Pacific University, Beppu, Oita 874-8577, Japan

[3] Sungkyunkwan University (SKKU), Suwon 16419, Korea

* Corresponding: hoang.nguyenminh@phenikaa-uni.edu.vn

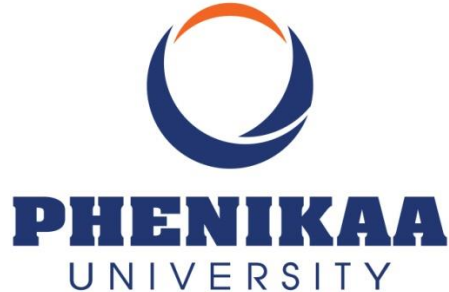

(Un-peer-reviewed manuscript v3)

August 01, 2021

---

Research manuscript to celebrate the 4th anniversary of the Centre for Interdisciplinary Social Research (ISR), of Hanoi-based Phenikaa University.

August 01, 2017 – August 01, 2021

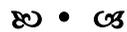

**Abstract**

Rapid urbanization with poor city planning has resulted in severe air pollution in low- and middle-income countries' urban areas. Given the adverse impacts of air pollution, many responses have been taken, including migration to another city. The current study explores the psychological process and demographic predictors of migration intention among urban people in Hanoi, Vietnam – one of the most polluted capital cities in the world. The Bayesian Mindsponge Framework (BMF) was used to construct the model and perform Bayesian analysis on a stratified random sampling dataset of 475 urban people. We found that the migration intention was negatively associated with the individual's satisfaction with air quality. The association was moderated by the perceived availability of a better alternative (or nearby city with better air quality). However, the high migration cost due to geographical distance made the moderation effect of the perceived availability of a better alternative negligible. Moreover, it was also found that male and young people were more likely to migrate, but the brain drain hypothesis was not validated. The results hint that without air pollution mitigation measures, the dislocation of economic forces might occur and hinder sustainable urban development. Therefore, collaborative actions among levels of government, with the semi-conducting principle at heart, are recommended to reduce air pollution.

1. Introduction

In recent decades, rapid urbanization with poor city planning has caused air pollution to become a serious problem affecting many people in cities worldwide. The World Health Organization (WHO) estimates that air pollution is responsible for about seven million total deaths yearly, with 4.2 million deaths due to ambient air pollution. Low- and middle-income countries suffer the highest level of exposure [1]. During the COVID-19 pandemic, exposure to air pollution increases the risk of COVID-19 mortality [2,3]. For confronting air pollution, people have taken many adaptive strategies, such as reducing outdoor activities [4], increasing pharmaceutical purchases and medication usage [5], and particulate-filtering wearing facemasks [6]. Besides these responses, migration to another city is also a potential alternative when air pollution is inevitable. The current study's purpose, thus, is to explore the psychological process and demographic predictors of migration intention of urban people using a dataset from Hanoi, Vietnam – one of the most polluted capital cities in the world [7].

Air pollution does not only directly damage people's health, but it also has negative effects in various psychological, economic, and social aspects [8]. Regarding mental health, exposure to air pollution was found to associate with general psychological distress [9], depressive disorder [10,11], and even suicide [12-14]. It is also worth noting that the subjective perception of air pollution (with corresponding health risk) can also induce annoyance and health symptoms even at non-toxic exposure levels [15]. Regarding impacts on human cognition, studies have shown that urban air pollution has negative effects on children's cognitive development [16,17] and their academic performance [18]. Not only affecting the young, but air pollution also impairs cognitive functions in adults [19,20] and especially in the elderly, which was a risk for developing dementia [21,22]. Studies have consistently found that air pollution reduces productivity in physical laborers [23] and white-collar workers [24].

In Vietnam, the negativity of air pollution on urban people's health is critical. Luong, *et al.* [25], using the data on daily admissions from Vietnam National Hospital of Pediatrics and daily records of air pollutions, find that the increasing levels of $PM_{10}$, $PM_{2.5}$ or $PM_1$ were positively associated with respiratory admissions of young children in Hanoi. Also, exposure to smaller PM could lead to higher risk [25]. Moreover, air pollution is positively associated with cardiorespiratory hospitalizations [26] and pneumonia-related hospitalizations [27]. One of the main air pollution sources in Vietnam's urban areas is traffic [28]. Assessment of the health risk induced by mobility in Hanoi shows that 3,200 deaths could result from $PM_{10}$ emitted by traffic [29].

Given the adverse impacts of air pollution on human development, health, and productivity, many responses have been implemented across levels. At the international level, the United Nations has made several targets to reduce air pollution. For example, Target (3.9) aims to "substantially reduce the number of deaths and illnesses from hazardous chemicals and air", while Target (7.a) seeks to promote clean energy technology. At the national level, countries like China [30], European countries [31], and the United States [32], have enforced air pollution prevention and control policies and legislations.

At the individual level, to protect themselves from the health risks of air pollution, people living in polluted urban areas may have the defensive response of averting behavior – meaning they will limit exposure by lessening their time being outdoor [33]. As examples of this coping strategy, higher air pollution level was found to associate with a higher number of school absences [34], lower level of outdoor cycling activities [35], and less usage of public parks [36].

Beyond just avoiding spending time outdoor, environmental stress can also lead to migration as a response [37]. On national scales, internal migrations from provinces or regions with worse air quality to those with higher air quality were observed, such as in Iran [38], Italy [39], and China[40]. High levels of air pollution also negatively affect the migration rate [41] and migrants' desire to stay [42], leading to further loss of human capital in the local region.

It is worth noting that in China, researchers found the "brain drain" effect (outmigration of talents) associated with air pollution, which includes the loss of skilled workers [43,44] and college graduates [45]. Regarding international migration influenced by air pollution, a higher emigration rate exists among higher educated people [46]. Additionally, emigration interest associated with air pollution level is seen through the degree of migration-related information-seeking activities [47].

While past studies have provided many valuable findings on the relationship between air pollution and migration behavior, certain aspects acquire limited attention. Most studies have focused on macro-scale investigations that employ particulate matter measurements and recorded population flows and changes [38-40]. This approach faces difficulties when delving deeper into the psychological processes among migrants. Instead, using individual-scale data focusing on people's subjective perception of air pollution level and migration desire, the current study can help generate more insights about the psychological responses against this type of environmental stressor. It is worth noting that the current study aims to examine the ideation of migration (thoughts, desires, intentions), as opposed to completed actions.

Additionally, many studies on this topic have been conducted using data in Iran, Italy, the United States, and especially China, but little is known about the situation in other developing countries with a high level of air pollution. Here, we attempt to use the mindsponge information-processing framework to examine the psychological process of domestic migration intention among urban people in Vietnam. Although there are several studies about air pollution's adversities in Vietnam, air pollution-induced migration has been underresearched.

The psychological process includes three main factors: 1) satisfaction with current air quality, 2) perceived option availability, 3) and consideration of moving cost. Why these factors are considered would be explained in the subsection of model construction. Thus, one of the main research objectives of this study is:

- Examine the impacts of satisfaction with current air quality, perceived option availability, and consideration of moving cost on the migration intention.

Besides, for identifying the potential group of migrants due to air pollution, we also:

- Examine the associations between demographic characteristics and migration intention due to air pollution.

## 2. Materials and method

### 2.1. Materials

The data used in this study were retrieved from two open datasets about the perceptions towards air pollution among inhabitants of Hanoi [48,49]. These datasets are the results of two survey collections using stratified random sampling methods at the central and suburban areas of the city, respectively. The survey collections were conducted during November and December 2019. Hanoi was chosen as the study site due to three reasons: 1) Hanoi was ranked 7th among the most polluted capital cities around the world [7]; 2) Hanoi is a fast-growing city in Vietnam; 3) Hanoi is the second largest and most populous city in Vietnam.

According to Khuc, Phu and Luu [49], the survey collection consisted of three steps. First, the collectors were recruited and paid generously to encourage them to perform well during the collection process. The researchers also held two four-hour seminars to help the collectors understand the project's goals and the questionnaire. During the seminar, necessary skills and techniques to extract information from respondents are also instructed. Then, two pilot tests were conducted to ensure the final version is error-free, straightforward, and easy to be understood. Lastly, the collectors conducted face-to-face interviews with the respondents and maintained mutual interaction and communication to solve issues or questions arising during the survey collection.

| Variable | Meaning | Type of variable | Value |
|---|---|---|---|
| *MoveCity* | Whether the respondent had the intention to immigrate to other provinces due to air pollution | Binary | Yes = 1  No = 0 |
| *AirSatisfaction* | The respondent's satisfaction level towards the current air quality | Continuous | From 1 (very dissatisfied) to 5 (very satisfied) |
| *BetterCloseCity* | Whether the respondent perceived that neighboring provinces had better air quality | Binary | Yes = 1  No = 0 |

| *BetterSouthCity* | Whether the respondent perceived that Southern provinces had better air quality | Binary | Yes = 1<br>No = 0 |
|---|---|---|---|
| *AgeGroup* | Age group | Continuous | From 10 to 18 = 1<br>From 19 to 30 = 2<br>From 31 to 40 = 3<br>From 41 to 50 = 4<br>From 51 to 60 = 5<br>Above 60 = 6 |
| *Gender* | Gender | Binary | Male = 1<br>Female = 0 |
| *Education* | The highest level of education | Continuous | Secondary school or below = 1<br>Highschool = 2<br>Technical school, college degree, university degree = 3<br>Master degree = 4<br>Doctoral degree = 5 |

We generated seven variables for the statistical analysis from the dataset: six predictor variables and one outcome variable. The outcome variable is *MoveCity*, created from the question "Do you intend to move your family and work in another province with less pollution?" and two corresponding 'yes' and 'no' answers. The urban people's satisfaction with the current air quality level is determined by asking the question "Overall, how satisfied are you with the air quality?" and demonstrated by the *AirSatisfaction* variable. The air satisfaction level was rated based on a 4-point Likert scale ranging from 1 ('very dissatisfied') to 4 ('very satisfied').

*BetterCloseCity* and *BetterSouthCity* variables were modified from two original items in the dataset. Originally, Khuc, Phu and Luu [49] asked the respondents, "How do you feel about the air quality in Hanoi compared to neighboring provinces/cities?" and "How do you feel about the air quality in Hanoi compared to southern provinces/cities?" and provided them with four answers: 'better than', 'same as', 'less than', and 'I don't know'. For investigating the impact of perceived option availability, modifications were made to turn them into binary variables, with 1 being 'less than', and 0 'better than' and 'same as'. The respondents that reported 'I don't know' were excluded from the analysis.

### 2.2. Model construction

For assessing the behavior of domestic migration induced by air pollution, the BMF (or the Bayesian Mindsponge analytical approach) was employed. Specifically, we constructed three models based on the mindsponge framework of information processing [50,51] and performing Bayesian analysis with the constructed models. The framework is effectively applied in investigating the psychological mechanisms underneath human behaviors [52-54].

The first model aims to examine the impacts of individuals' satisfaction towards the current air quality and the moderation effects of perceived availability of a better option, and the lower cost of moving on the migration intention. The second model examines the relationships between socio-demographic factors and the likelihood of having migration intention to identify population groups with a high possibility of migrating. Model 3 combines Model 1 and Model 2 to test the robustness of results in prior models.

The mindsponge mechanism suggests that the information must pass through a multi-filtering system using cost-benefit judgments for ideation to occur within an individual's mind. Here, we assume that an individual will intend to move and work in another city when the idea of migration is accepted into their mindset because it is considered a beneficial option. On the other hand, if migration is considered disadvantageous, it will be rejected, and the individual will not have the intention to migrate.

Based on this assumption, we assumed that an individual with a higher level of satisfaction towards the current air quality perceives the act of migration to another city as less beneficial. Moreover, we suspected the existence of two other factors that influence the cost-benefit judgment of the migration idea by moderating the air quality satisfaction's impact on the migration intention, namely: 1) the perceived availability of a better option, and 2) the lower perceived cost of moving. Thus, we postulated that the migration intention is stronger if people notice other places with better living conditions and lower moving costs.

For estimating the moderation effect of perceived availability of better an option, we added two variables, *BetterCloseCity* and *BetterSouthCity*, into the model as interactions with the variable *AirSatisfaction*. The impact of the perceived moving cost will be justified by the difference between the effects of *AirSatisfaction\*BetterCloseCity* and *AirSatisfaction\*BetterSouthCity*. More specifically, if the effect of *AirSatisfaction\*BetterCloseCity* is larger than the effect of *AirSatisfaction\*BetterSouthCity*, the positive impact of lower moving cost on migration intention can be confirmed. Otherwise, the postulation would be rejected.

Model 1 can be presented by the following specification:

**Model 1:** $MoveCity \sim \alpha + AirSatisfaction + AirSatisfaction * BetterClosedCity + AirSatisfaction * BetterClosedCity$

In the second model, we focus on the relationships between the socio-demographic factors (age, gender, and education) and the migration intention for identifying the demographic characteristics of potential migrators, so the model is constructed as follows:

**Model 2:** $MoveCity \sim \alpha + Age + Gender + Education$

Model 3 is the combination of the first and second models, described as the equation below:

**Model 3:** $MoveCity \sim \alpha + Age + Gender + Education + AirSatisfaction + AirSatisfaction * BetterCloseCity + AirSatisfaction * BetterSouthCity$

### 2.3. Methods and validation

There are five reasons that Bayesian analysis was employed in the current study. First of all, science is now facing the reproducibility crisis that a large proportion of studies across disciplines could not be replicated. Psychology [55] and social sciences [56] are not excluded. One of the main reasons is suggested to be the wide sample-to-sample variability in the *p*-value. For not being dependent on the *p*-value, the Bayesian inference approach is a good alternative because it treats all the properties probabilistically, including the unknown parameters. Secondly, this characteristic of Bayesian analysis has high compatibility with the current study's design, which is explanatory research. The study employed the mindsponge framework to explain the psychological process that might lead to the migration intention. By treating all properties probabilistically, the Bayesian analysis helps us consider the impacts of other unknown factors while maintaining the rule of parsimony for the explanation [57].

Our models examined the effects of two variables: *BetterCloseCity* and *BetterSouthCity*. Regressing two seemingly correlational variables might result in confounding outcomes. Thus, we determined to treat those variables as interaction variables for turning their effects into non-linear. The treatment made the model more complex and require a larger sample size for sound estimation [58]. Nevertheless, the Markov Chain Monte Carlo method integrating the Bayesian analysis generates a large number of parameters' samples through stochastic processes of Markov chains, which helps fit complex models effectively. Moreover, thanks to the MCMC, Bayesian analysis could provide a more precise estimation for a small sample size dataset than the frequentist approach [59].

Finally, prior distribution incorporation is another advantage of Bayesian analysis. Even though we set priors as 'uninformative' to avoid the subjective influences over the simulated outcomes, the prior function still can be capitalized to check the robustness of the simulated results by performing the "prior-tweaking" technique.

For validating the simulated posterior outcomes, we adopt a four-pronged validation strategy. Initially, we conducted a goodness-of-fit check on any simulated model using the PSIS-LOO diagnostic plots [60]. If the *k* values shown on the plot are all below 0.5, the model can be deemed to have a good fit with the data. Next, we continued with the convergence check using both the diagnostic statistics and plots. The statistics include the effective sample size, and the Gelman shrink factor, while the plots include the trace plot, Gelman plot, and an autocorrelation plot. Then, we combined Model 1 and Model 2 into Model 3 to check the results' robustness. We also compare the weight among models during this stage to see their goodness-of-fit levels with the data. Finally, the prior-tweaking technique was performed. Further explanation and interpretation of the diagnostic statistics, plots, weight comparison, and prior-tweaking technique are presented in the Results section.

The **bayesvl** R package was used to perform Bayesian analysis in the current study for three reasons: 1) it is a cost-effective alternative; 2) it has a great capability to visualize eye-catching graphics; and 3) it is easy to be operated [61,62]. The dataset, data description, and code snippets of Bayesian analysis were deposited on the The Open Science Framework for later replications (https://osf.io/us5tr/).

## 3. Results

To test the assumptions we made above, we conducted Bayesian analysis to examine Model 1 and Model 2, while Model 3 was used to test the robustness of the results. For estimating these three models, we employed Bayesian MCMC simulation with 5,000 iterations, 2,000 warm-up iterations, and four Markov chains. All the models' simulated results and their technical validity using priors being 'uninformative' are presented accordingly. The analysis was performed on the samples with more proportion of male than female respondents (54.53% versus 45.26%). A majority of the samples belonged to the age group between 10 and 30 (61.47%). Among 475 respondents, approximately 5% reported their intention to move their family and work in another city due to air pollution.

### 3.1. Model 1: Migration cost-benefit judgment

The first model examined the effects of citizens' satisfaction with air quality and its interactions with perceived better air quality in neighboring and Southern cities on migration intention. The model's logical connection is shown in Figure 1.

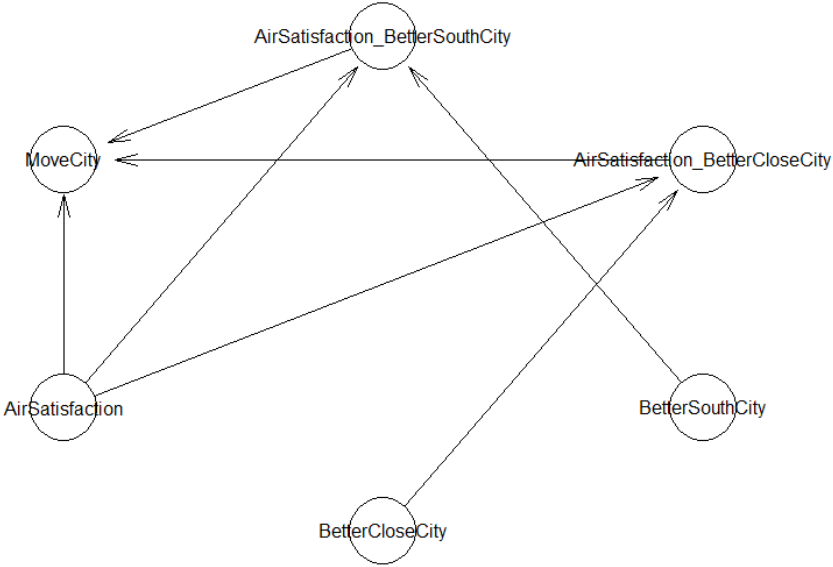

Figure 1: Model 1's logical network

The PSIS diagnostic plot shows that all *k* values are below 0.5, suggesting that Model 1 has a high goodness-of-fit with the data (see Figure 2).

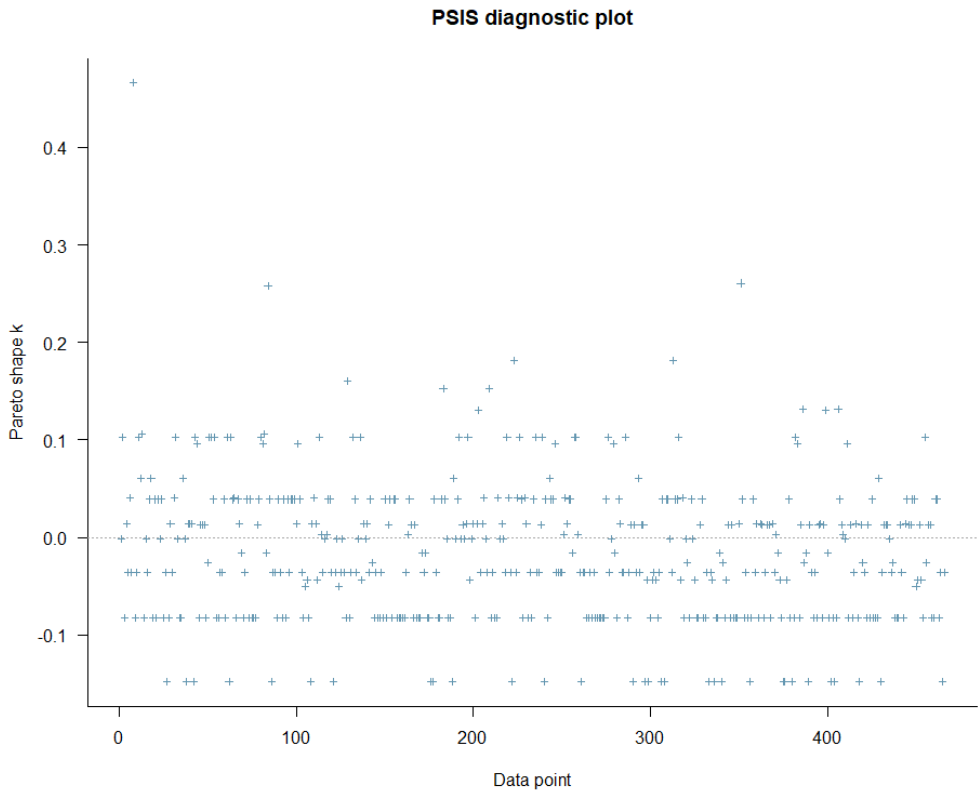

Figure 2: Model 1's PSIS diagnostic plot

The effective sample size (n_eff > 1000) and Gelman shrink factor (Rhat = 1) of all simulated posterior coefficients portray a good convergence of the model's Markov chains (see Table 2). The convergence is also visually diagnosed using the trace plots, autocorrelation plots, and Gelman plots.

**Table 2:** Model 1's simulated posteriors.

| Parameters | Uninformative | | Prior-tweaking (high reliability) | | Prior-tweaking (low reliability) | | n_eff | Rhat |
|---|---|---|---|---|---|---|---|---|
| | Mean | SD | Mean | SD | Mean | SD | | |
| Constant | -1.44 | 0.45 | -1.35 | 0.41 | -1.62 | 0.42 | 6397 | 1 |
| AirSatisfaction | -0.67 | 0.31 | -0.76 | 0.27 | -0.49 | 0.25 | 4424 | 1 |
| AirSatisfaction*BetterCloseCity | 0.26 | 0.25 | 0.31 | 0.24 | 0.16 | 0.21 | 6160 | 1 |
| AirSatisfaction*BetterSouthCity | 0.06 | 0.17 | 0.06 | 0.17 | 0.07 | 0.17 | 8012 | 1 |

Note:

*SD = Standard deviation*

*\*\* The effective sample size (n_eff) and Gelman value (Rhat) of simulated results with different priors are almost similar, so only the n_eff and Rhat of simulated results using uninformative priors are presented.*

Figure 3 demonstrates the trace plots of all posterior parameters. The y-axis of the trace plot represents the posterior values of each parameter, while the x-axis represents the iteration order of the simulation. The colored lines in the middle of the trace plot are Markov chains. If the Markov chains fluctuate around a central equilibrium, they can be considered good-mixing and stationary. These two characteristics are a good signal of convergence.

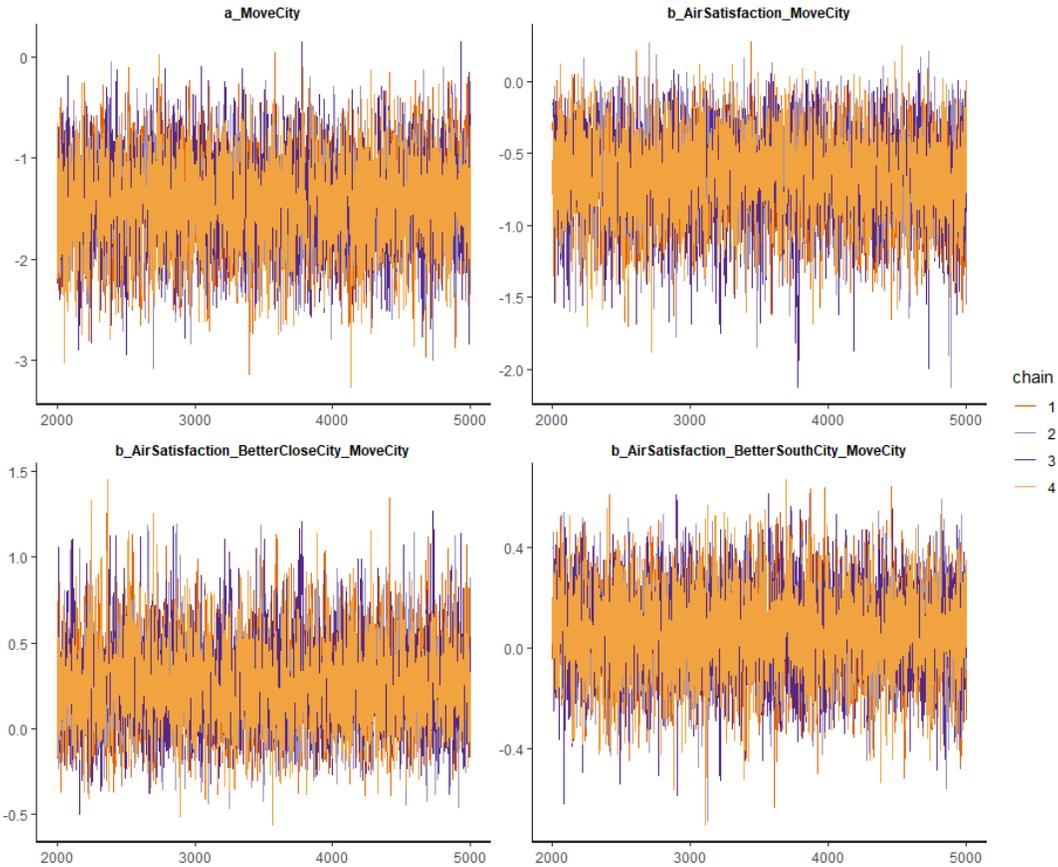

Figure 3: Trace plots for Model 1's posterior parameters

Gelman plots of Model 1's parameters are shown in Figure 4. The y-axis of the Gelman plot illustrates the shrink factor (or Gelman factor), which is used to estimate the relative between the variance between Markov chains and the variance within chains. Meanwhile, the x-axis demonstrates the iteration order of the simulation. As can be seen that, the shrink factors of all parameters drop rapidly to 1 during the warm-up iterations, hinting that there is no divergence among Markov chains. Therefore, the Markov property is held.

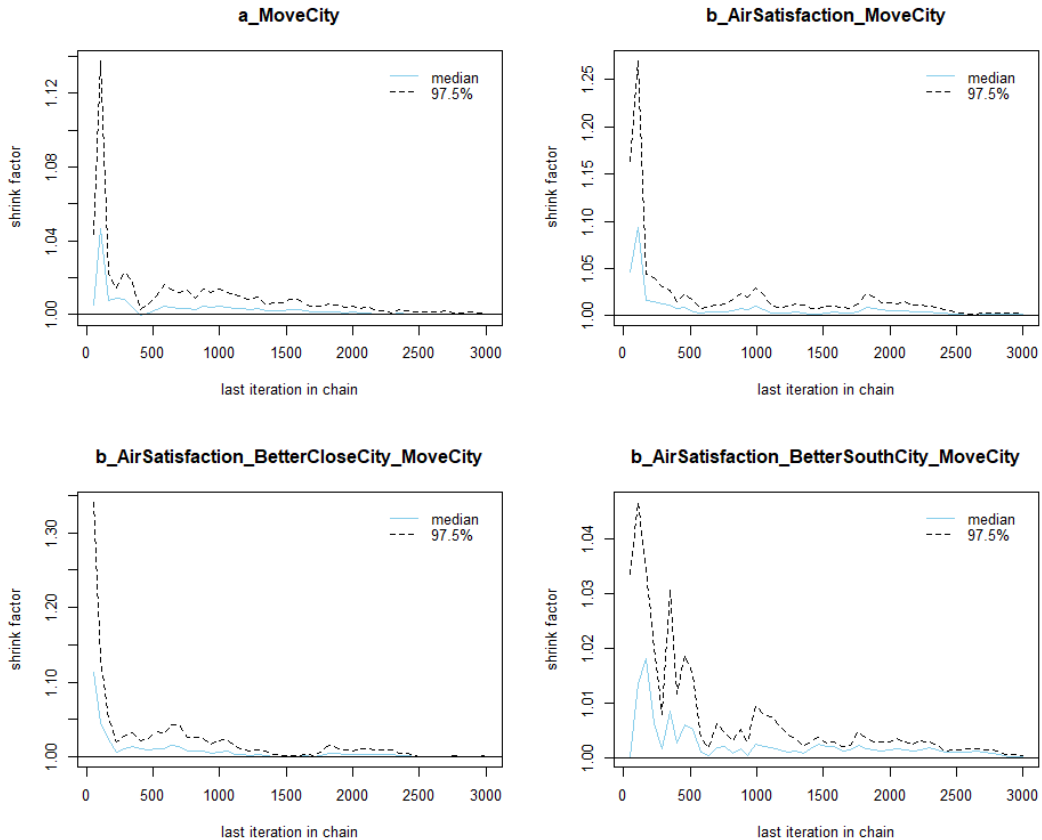

Figure 4: Gelman plots for Model 1's posterior parameters

Another further step to validate the convergence of Model 1's is to diagnose the Markov chains' autocorrelation levels visually (see Figure 5). The x-axes of the autocorrelation plots represent the number of Markov chains' lag, while the y-axes show the average level of autocorrelation of each chain. Visually, the average autocorrelation level declines substantially before the fifth lag, inducing all parameters to acquire a great number of effective samples. The autocorrelation plots' demonstrations again confirm the convergence of Model 1's Markov chains.

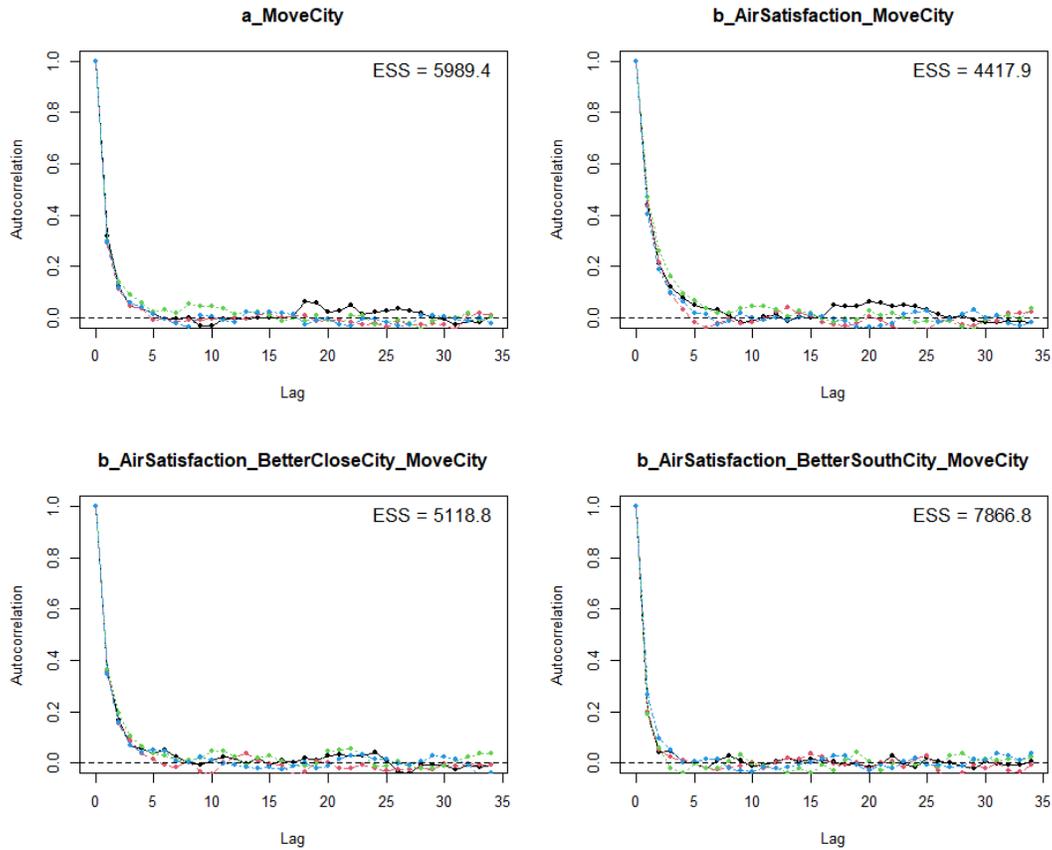

Figure 5: Autocorrelation plots for Model 1's posterior parameters

From the simulated posterior results of Model 1, we found that the citizens' satisfaction towards air quality was negatively associated with the intention to migrate to another city ($\mu_{AirSatisfaction} = -0.67$ and $\sigma_{AirSatisfaction} = 0.31$). This result confirms our assumption that more satisfied people are less likely to migrate due to the lower benefit of leaving the city. However, the cost-benefit judgment of urban citizens about migration due to air pollution is much more complex. Perceived living places with better air quality (either neighboring provinces or further provinces in the South) moderated the effect of air satisfaction on migration intention ($\mu_{AirSatisfaction*BetterCloseCity} = 0.26$ and $\sigma_{AirSatisfaction*BetterCloseCity} = 0.25$; $\mu_{AirSatisfaction*BetterSouthCity} = 0.06$ and $\sigma_{AirSatisfaction*BetterSouthCity} = 0.17$). The moderation impact of the perceived availability of a better alternative in the South is smaller than the alternative nearby. These results validate our assumptions on the moderation effects of the perceived availability of a better option and the lower cost of moving, respectively.

For robustness check, prior-tweaking was performed. In both cases (high prior reliability and low prior reliability), the coefficients' effect patterns did not change, though the statistics slightly changed. We can conclude that the effects in Model 1 are robust even when the prior reliability varies.

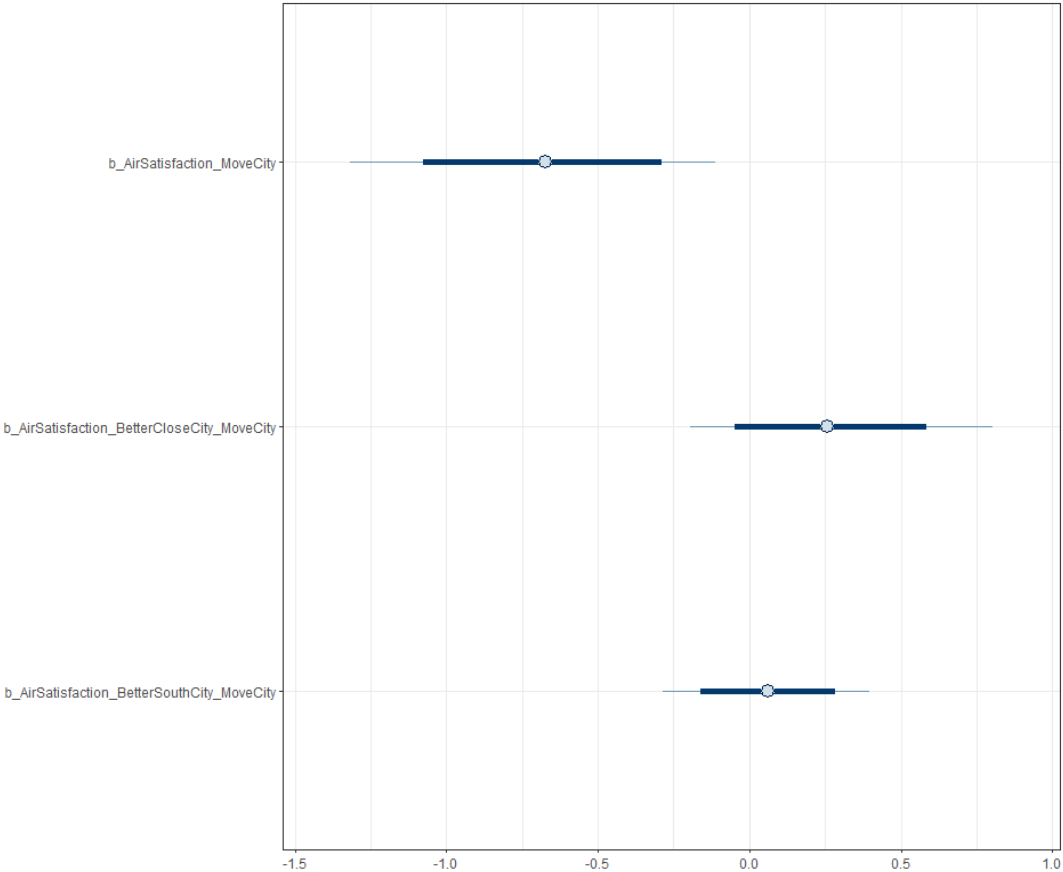

Figure 5: Distributions of Model 2b's posterior coefficients on an interval plot.

The distributions of Model 1's parameters are visualized in the interval plot (see Figures 5) for assessing their reliability. The posterior values of parameters are shown in the x-axis of the plot. The distribution of coefficient *AirSatisfaction* lies entirely on the negative side of the axis, indicating a highly reliable negative association between *AirSatisfaction* and *MoveCity*. Distributions of coefficients *AirSatisfaction\*BetterCloseCity* and *AirSatisfaction\*BetterSouthCity* are mostly located on the positive side, implying that the moderation effects of perceived availability of a better option and lower cost of moving had the highest probability to be positive. It is shown in Figure 6 that *AirSatisfaction\*BetterCloseCity* has a greater moderation effect. A higher proportion of simulated values have positive values according to the x-axis than y-axis, so the moderation effect of *AirSatisfaction\*BetterCloseCity* can be deemed to be more reliable than *AirSatisfaction\*BetterSouthCity*.

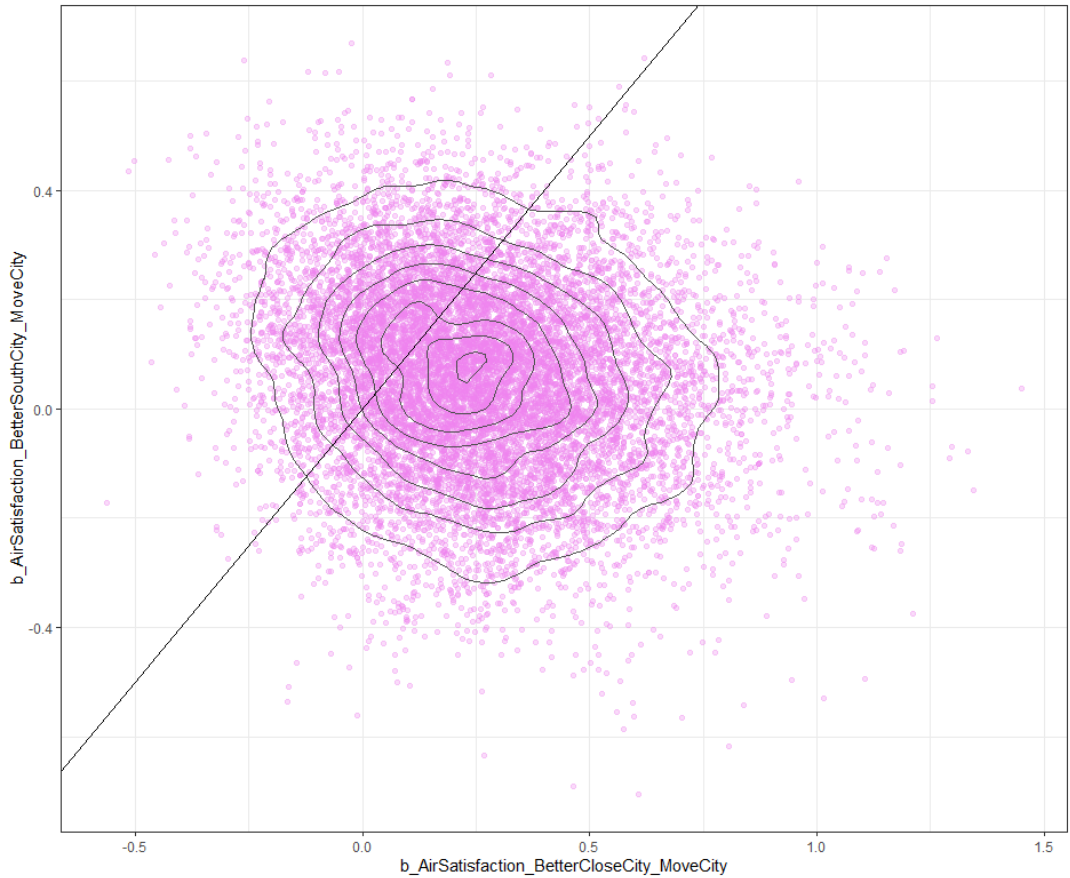

Figure 6: Pairwise distribution plot for Model 1's *AirSatisfaction\*BetterCloseCity* and *AirSatisfaction\*BetterSouthCity*

### *3.2.   Model 2: Socio-demographic factors*

In the second model, we examine the impacts of age (*Age*), gender (*Gender*) and education (*Education*) on migration intention. The logical network of Model 2 can be illustrated in Figure 7.

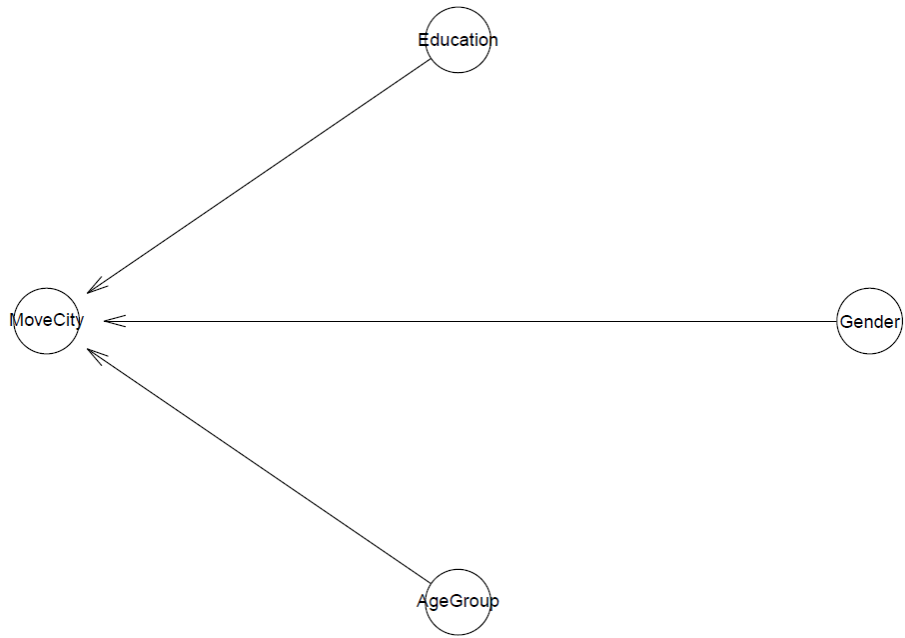

Figure 7: Model 2's logical network

The model's goodness-of-fit is relatively high as there are no *k* values on the PSIS diagnostic plot higher than 0.5 (see Figure 8).

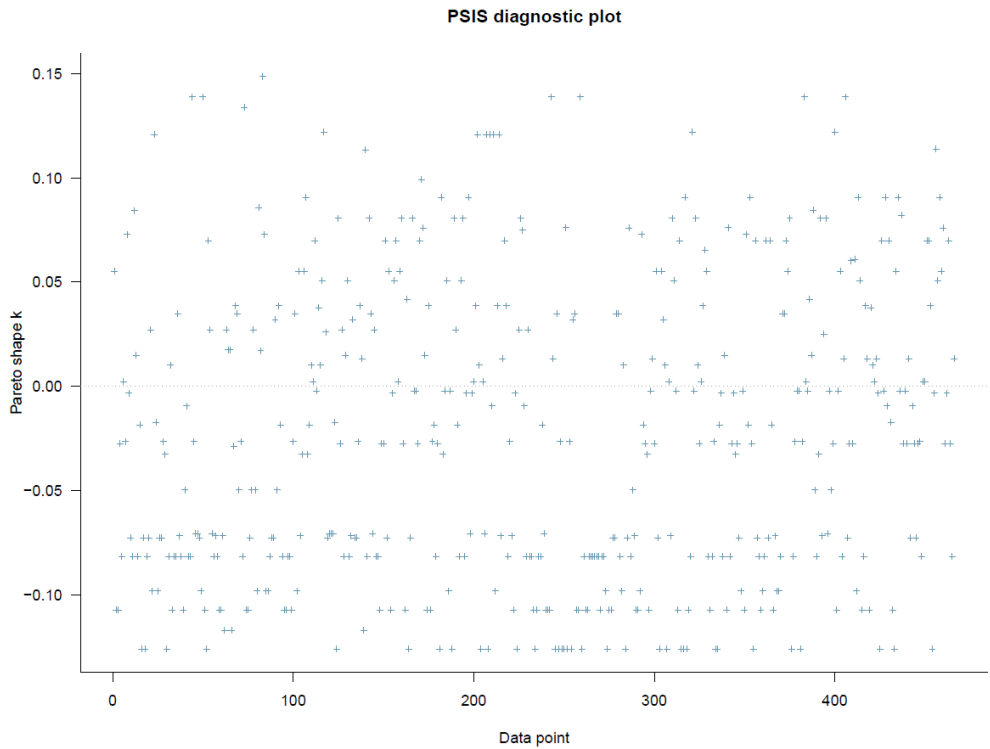

Figure 8: Model 2's PSIS diagnostic plot

The diagnostic statistics of Model 2 indicate that the posterior results are well convergent. More specifically, the n_eff values are larger than 1,000, while the Rhat values are equal to 1. The visual diagnostic methods, like the trace plots (see Figure A1), the Gelman plots (see Figure A2), and the autocorrelation plots (see Figure A3), also confirm Model 2's Markov chains' convergence.

Table 2: Model 2's simulated posterior coefficients

| Parameters | Uninformative | | Prior-tweaking (high reliability) | | Prior-tweaking (low reliability) | | n_eff** | Rhat** |
|---|---|---|---|---|---|---|---|---|
| | Mean | SD* | Mean | SD | Mean | SD | | |
| Constant | -1.54 | 0.89 | -1.33 | 0.88 | -1.61 | 0.87 | 4314 | 1 |
| AgeGroup | -0.31 | 0.13 | -0.35 | 0.13 | -0.29 | 0.12 | 5513 | 1 |
| Gender | 0.59 | 0.34 | 0.58 | 0.35 | 0.57 | 0.34 | 7542 | 1 |
| Education | -0.02 | 0.24 | -0.06 | 0.24 | -0.02 | 0.24 | 5176 | 1 |

Note:

*SD = Standard deviation*

*\*\* The effective sample size (n_eff) and Gelman value (Rhat) of simulated results with different priors are almost similar, so only the n_eff and Rhat of simulated results using uninformative priors are presented.*

The simulated posterior results show that *AgeGroup* and *Gender* have clear impacts on the migration intention, whereas the effect of *Education* is minimal and can be neglected ($\mu_{Education} = 0.59$ and $\sigma_{Education} = 0.34$). *AgeGroup* variable is found to be negatively associated with *MoveCity* variable ($\mu_{AgeGroup} = -0.31$ and $\sigma_{AgeGroup} = 0.13$). To elaborate, urban citizens who belonged to older age groups had less probability of migrating to other provinces due to air pollution. The association between *Gender* and *MoveCity* variables is negative ($\mu_{Gender} = 0.59$ and $\sigma_{Gender} = 0.34$), implying that male citizens were more likely to migrate to other provinces due to air pollution than female citizens.

The distributions of Model 2's parameters are shown in Figure 9. The bold black lines at the feet of the histograms represent the highest posterior distribution interval (HPDI) at 89% of the parameters. As can be seen that the HPDIs of *AgeGroup* and *Gender* variables are located completely on the negative and positive sides of the x-axis's origin, so the effects of citizen's age and gender on moving intention are reliable. Even when being simulated using priors indicating different reliability levels, the effects of age and gender only change slightly.

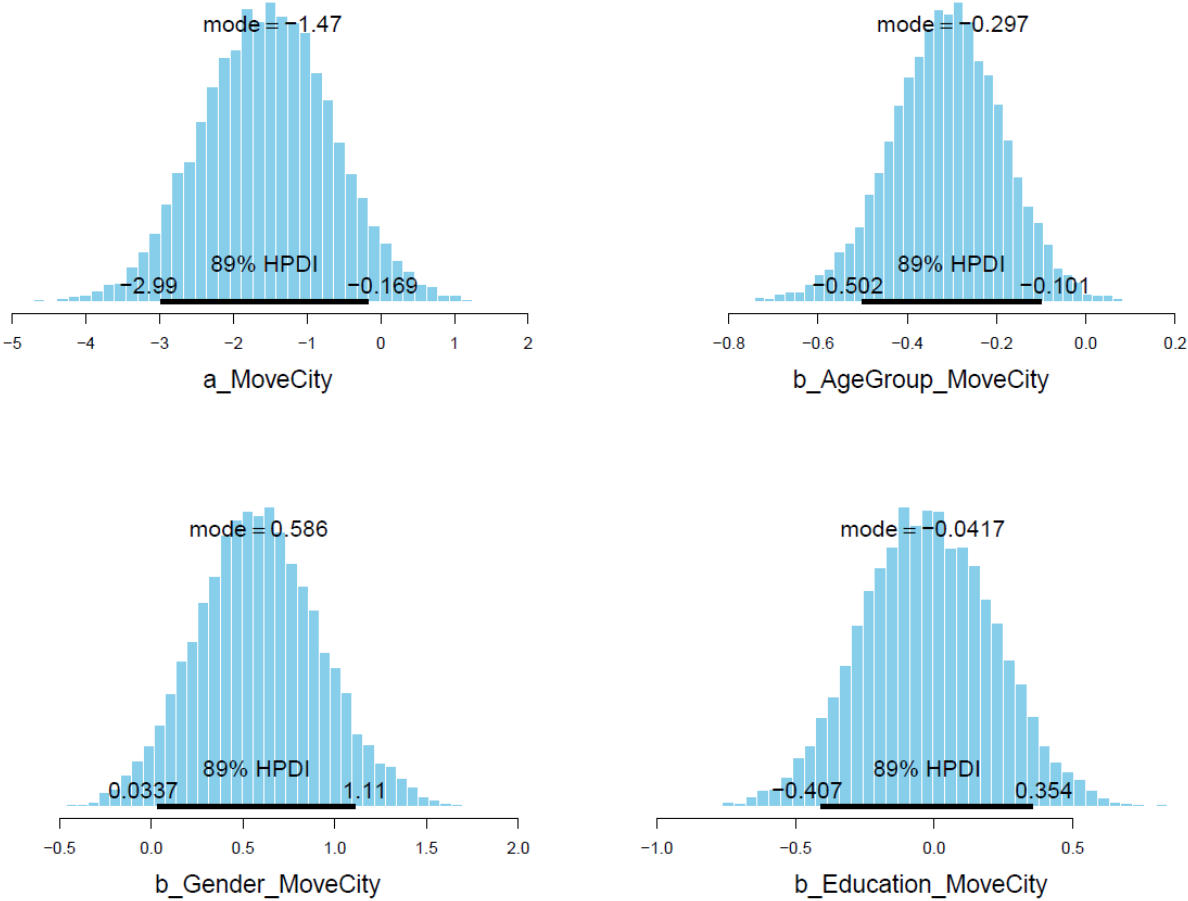

Figure 9: Distributions Model 2a's posterior coefficients with HPDI 89%

### 3.3. Model 3: Robustness check and model comparison

The third model combines Models 1 and 2 to test the robustness of the results presented above and the models' goodness-of-fit. Figure 10 illustrates the logical network of Model 3.

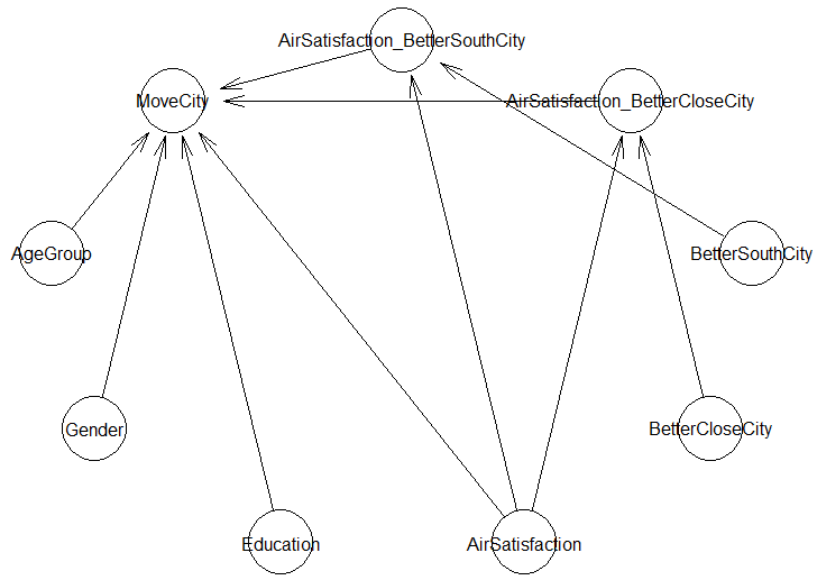

Figure 10: Model 3's logical network

Despite combining Models 1 and 2 into Model 3, the PSIS diagnostic plot of Model 3 still indicates that the model has a high goodness-of-fit with the data ($k < 0.5$).

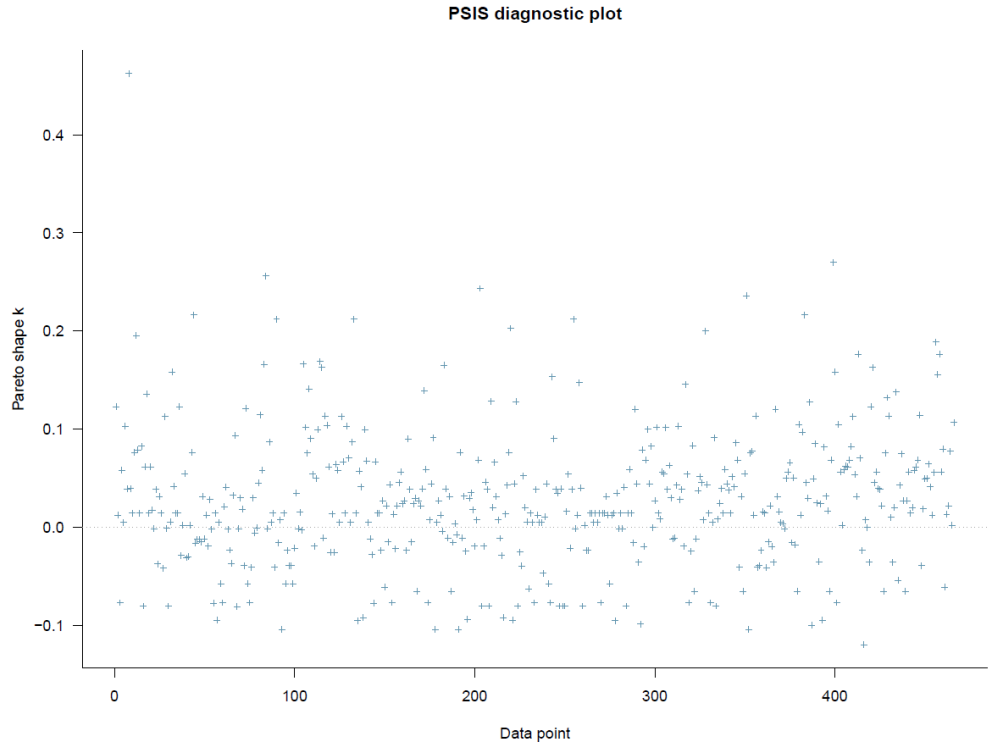

**Figure 10:** Model 3's PSIS diagnostic plot.

Visually, the trace plots for Model 3's posterior parameters show "clean and healthy" Markov chains (see Figure A4). Moreover, the Gelman and autocorrelation plots (see Figures A5 and A6) also demonstrate a good convergence signal. As a result, the Markov chain central limit theorem holds in Model 2's simulation. The diagnostic statistics confirm this statement, as all the n_eff values are greater than 1000, and Rhat values are equal to 1.

**Table 3:** Model 2c's simulated posterior coefficients

| Parameters | Uninformative | | Prior-tweaking (high reliability) | | Prior-tweaking (low reliability) | | n_eff | Rhat |
|---|---|---|---|---|---|---|---|---|
| | Mean | SD* | Mean | SD | Mean | SD | | |
| Constant | -0.17 | 1.06 | -0.10 | 1.03 | -0.43 | 1.04 | 7921 | 1 |
| AgeGroup | -0.35 | 0.13 | -0.34 | 0.13 | -0.34 | 0.13 | 8471 | 1 |
| Gender | 0.61 | 0.34 | 0.61 | 0.34 | 0.60 | 0.34 | 10351 | 1 |
| Education | -0.12 | 0.25 | -0.12 | 0.25 | -0.09 | 0.25 | 6051 | 1 |
| AirSatisfaction | -0.71 | 0.32 | -0.78 | 0.27 | -0.51 | 0.25 | 7561 | 1 |
| AirSatisfaction*BetterCloseCity | 0.21 | 0.25 | 0.25 | 0.25 | 0.10 | 0.22 | 7912 | 1 |
| AirSatisfaction*BetterSouthCity | 0.00 | 0.18 | -0.01 | 0.18 | 0.00 | 0.17 | 11571 | 1 |

The effects of all coefficients simulated in Model 3 are consistent with those in Models 1 and 2. In detail, *AgeGroup* and *AirSatisfaction* are negatively associated with *MoveCity* ($\mu_{AgeGroup} = -0.35$ and $\sigma_{AgeGroup} = 0.13$; $\mu_{AirSatisfaction} = -0.71$ and $\sigma_{AirSatisfaction} = 0.32$). The association between *Gender* and *MoveCity* remains positive ($\mu_{Gender} = 0.61$ and $\sigma_{Gender} = 0.34$). Even though the moderation effect of *BetterSouthCity* on the relationship between *AirSatisfaction* and *MoveCity* is negligible ($\mu_{AirSatisfaction*BetterSouthCity} = 0.00$ and $\sigma_{AirSatisfaction*BetterSouthCity} = 0.18$), it validates our assumption on the impact of moving cost on the citizens' migration cost-benefit judgment because *BetterCloseCity* still have a positive association with *MoveCity* ($\mu_{AirSatisfaction*BetterCloseCity} = 0.21$ and $\sigma_{AirSatisfaction*BetterCloseCity} = 0.25$). Using the prior-tweaking techniques, we also found no significant changes in the simulated posterior results. All coefficients' distributions are presented in Figure 11.

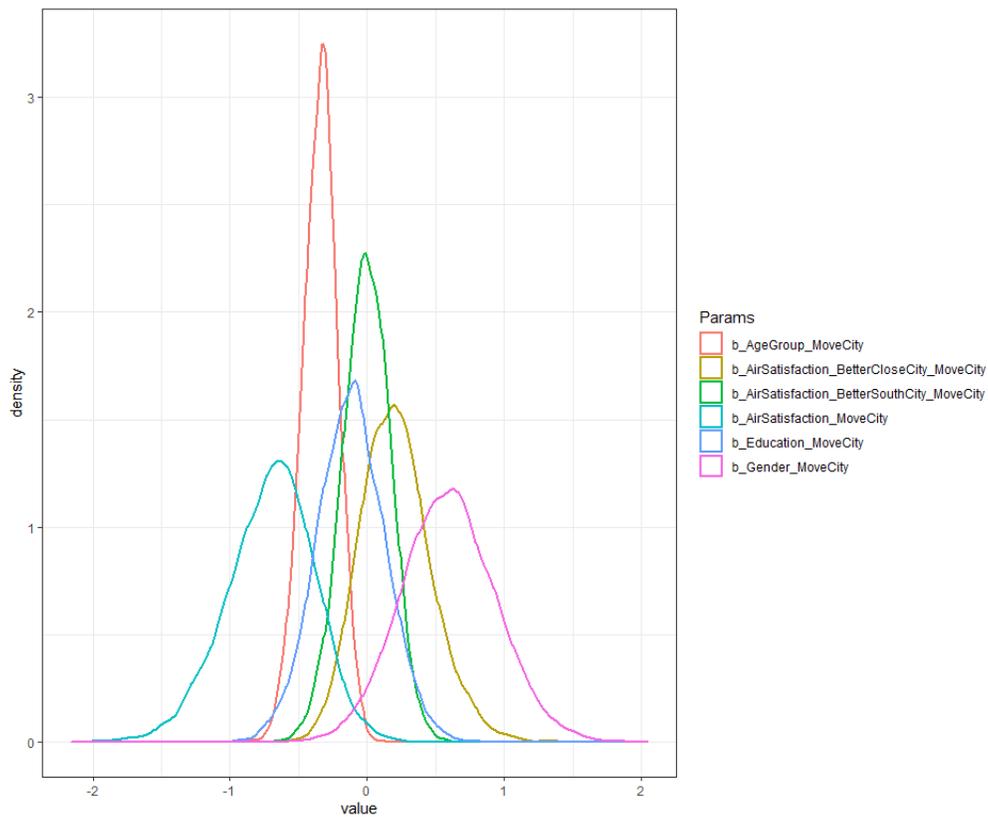

Figure 11: Distributions of Model 3's posterior coefficients on a density plot.

We employed four different values for comparing Models 1-3: WAIC, Pseudo-BMA without Bayesian bootstrap, Pseudo-BMA with Bayesian bootstrap, and Bayesian stacking (see Table 4). The Akaike weight was also used to rescale these values [63]. A total weight of 1 is partitioned among the three models, with the greater weight model implying better predictive accuracy on the data. Model 2, using socio-demographic factors as predictors, has the highest weight in all categories. While Model 3 acquires considerable weight in terms of WAIC (0.307), Pseudo-BMA without Bayesian bootstrap (0.294), and Pseudo-BMA with Bayesian bootstrap (0.285), Model 1 constitutes 0.214 and 0.383 of weight in the Pseudo-BMA with Bayesian bootstrap and Bayesian stacking categories, respectively. Given the considerable weight holding by each model, any of Models 1-3 has adequate predictive accuracy on the data. Thus, posterior results generated in this study can be deemed reliable, regardless of model (separate or combined).

Table 4: Weight comparison

|  | WAIC | Pseudo-BMA without Bayesian bootstrap | Pseudo-BMA with Bayesian bootstrap | Bayesian stacking |
|---|---|---|---|---|
| Model 1 | 0.042 | 0.041 | 0.214 | 0.383 |
| Model 2 | 0.650 | 0.665 | 0.501 | 0.544 |
| Model 3 | 0.307 | 0.294 | 0.285 | 0.073 |

## 4. Discussion

The current study is one of the first studies that examine the psychological perspectives of urban people on migration due to air pollution. It employed the BMF appoach on the responses of 475 urban people participating in a stratified random sampling survey collection in Hanoi – one of the most polluted capital cities in the world. Several major findings were identified from the simulated posterior results.

First, our results show that the intention to migrate to another living place is negatively associated with the satisfaction with the current environment's air quality (see Model 1). In other words, people who dislike the polluted air in their city are more likely to have an intention to move away. This is consistent with findings in prior studies using macro-indicators investigating the relationship between air pollution and internal migration behavior [38-40]. Our results support the idea that people migrate away from their current living area with high pollution levels as an averting strategy to reduce environmental health risks [37]. This specific type of deliberate avoidance behavior can be elaborated in terms of information processing based on the mindsponge framework's principles as follows.

Specifically, there are two fundamental conditions for a value to be accepted into the mindset: information availability/accessibility and a positive evaluation of the cost-benefit judgment. Thus, the ideation of migration to another city can be deemed to appear in an individual's mindset when an individual perceives it to be beneficial. Intuitively, the information about environmental health risks induced by poor air quality holds clear negative value in the cost-benefit scale (in other words, living in a polluted place is subjectively perceived as "costly"). This, in turn, adds more benefit to migration and increases the possibility of the migration idea being accepted into the mindset.

The cost-benefit judgment process of whether to move or not is multiplex and dependent on many factors, including but not limited to the perceived availability of a better alternative and cost of migration. We found that while the perceived availability of a better nearby alternative moderated the effect of air satisfaction on the migration intention, the perceived availability of a better alternative in the South did not. The perceived cost caused by geographical distance might be largely attributable to this finding. Besides financial cost, increasing psychic cost and information-diminishing might contribute to the negative effect of long distance on migration intention [64,65].

Demographic factors are also important predictors of migration intention. Our results showed that younger people (compared to older ones) and males (compared to females) were more likely to have migration intention due to air pollution. This might be because older people generally have a lower capacity (also self-perceived capacity) to relocate due to obstacles such as financial situation (e.g., retirement), health issues (e.g., physical limitations of old age), and adaptability to new living environments (e.g., high attachment to familiarities). Regarding gender, Vietnamese women often prioritize taking care of children and other family members, and these connections may cause the decision of migrating to be more costly for females. Nevertheless, we should not firmly overgeneralize the influences of age and gender on migration behaviors because they are very complex and situation-specific [66].

Regarding the effect of educational level, we found that there is no significant influence. This result is not consistent with the brain drain hypothesis in Chinese urban areas, suggesting that people with higher educational levels were more likely to migrate away from the polluted environment, leading to the local loss of high-quality human resources [43-45]. The inconsistency might be due to country/region-specific situations and differences in study samples' demographics (physical workers, office workers, and college graduates in former Chinese studies compared to the general urban population in our present study). The interactions between unexamined factors may be quite complex. For example, on the one hand, people with higher educational levels usually have higher pay and better working conditions, so the opportunity cost of leaving is higher. On the other hand, people with lower educational levels may perceive leaving the current job as risky as there is a possibility that they could not find a job in a new place. Furthermore, the inconsistency can be attributed to the differences between ideation (as examined in our present study) and completed behavior (as examined in prior studies in China).

The psychological process of migration intention indicates that domestic migration could be highly probable, especially among young and male populations. If the air pollution is not reduced, public awareness towards air pollution will rise [67] and affect the migration choice [42]. Eventually, the domestic migration induced by poor air quality might lead to the relocation of economic forces (male and young people) and hinder the sustainable development in the origin city.

Thus, collaborative actions among levels of government are required to mitigate air pollution from the main sources of urban air pollution in Vietnam, like traffic, industrial emissions, and construction sites [28,68]. More specifically, the national government should provide legislative and financial support to the municipal government. In contrast, the municipal government has to take the responsibility to provide support and enforce strict environmental regulations towards the private sector and improve urban planning quality and public communication regarding air pollution [69,70]. For air pollution to be effectively and efficiently reduced, the semi-conducting principle, suggesting that monetary values cannot be used interchangeably for environmental values, should be employed as the core ideology when designing, planning, and implementing mitigation measures [71].

Limitations of the current study are described here for transparency [72]. The current study only employed data from Hanoi capital city, so the findings regarding associations between demographic factors and migration intention might be restricted to the socio-cultural characteristics of the study site. However, as data of this study were gathered from a stratified random sampling survey collection, and the cost-benefit judgment mentioned above is relatively universal, findings on the psychological process of migration intention can be generalized. Moreover, there is a certain gap between the ideation and behavior, so the migration intention in our study should be interpreted as a risk rather than actual action.

# Appendix

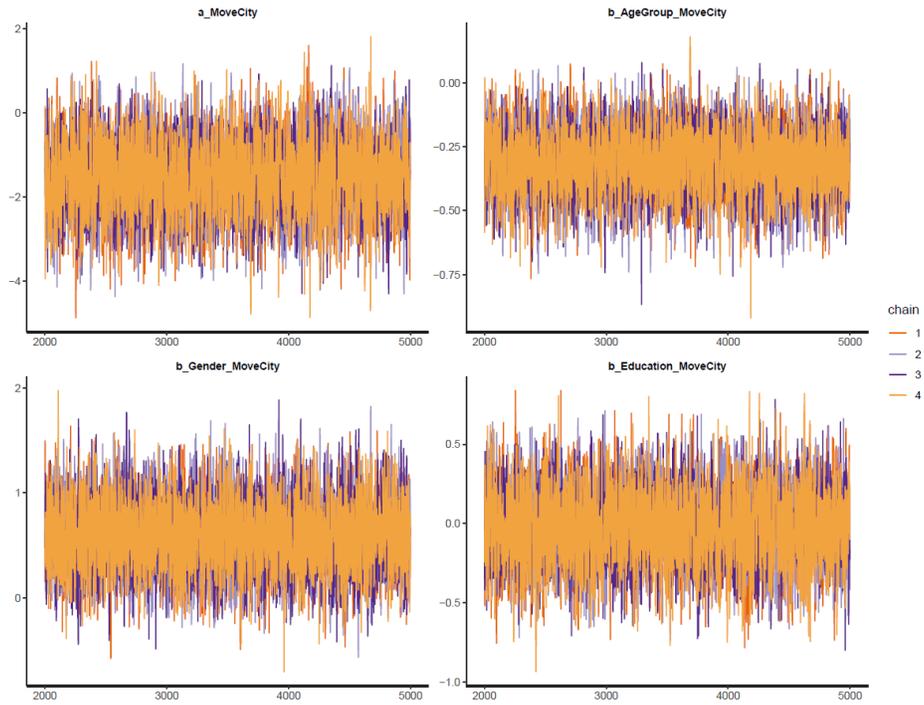

Figure A1: Trace plots for Model 2's posterior parameters

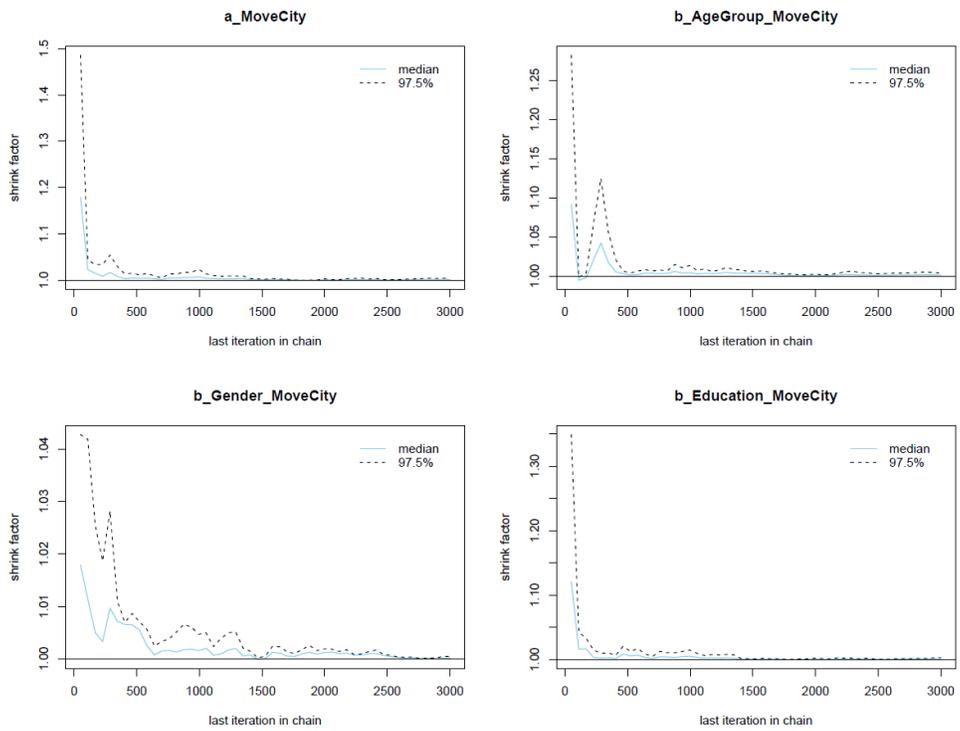

Figure A2: Gelman plots for Model 2's posterior parameters

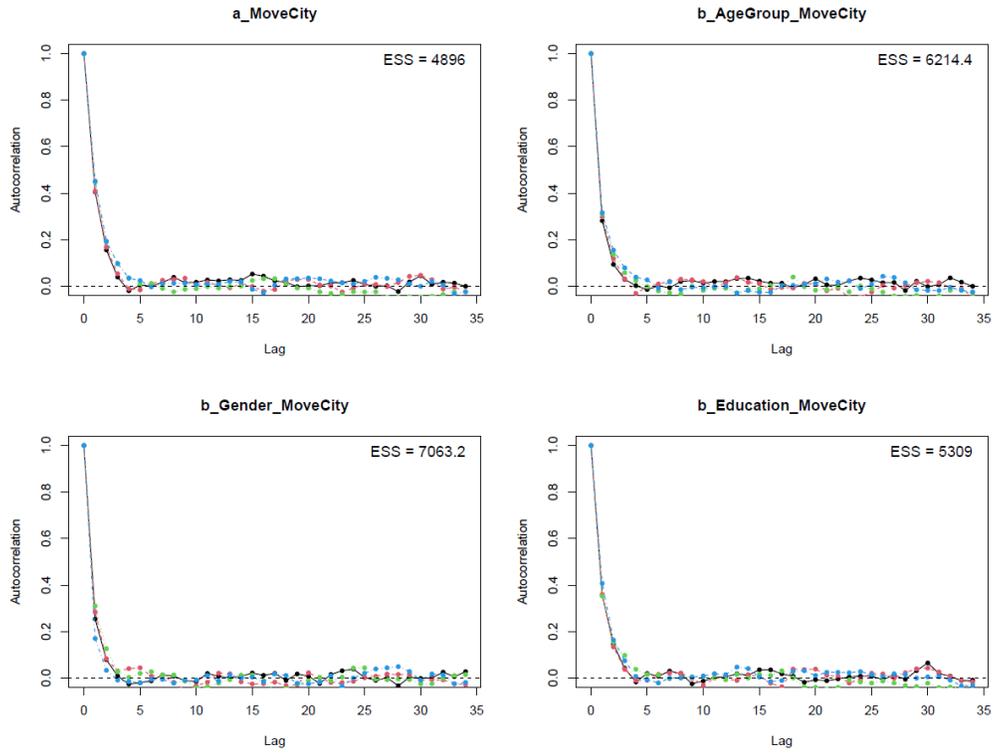

Figure A3: Autocorrelation plots for Model 2's posterior parameters

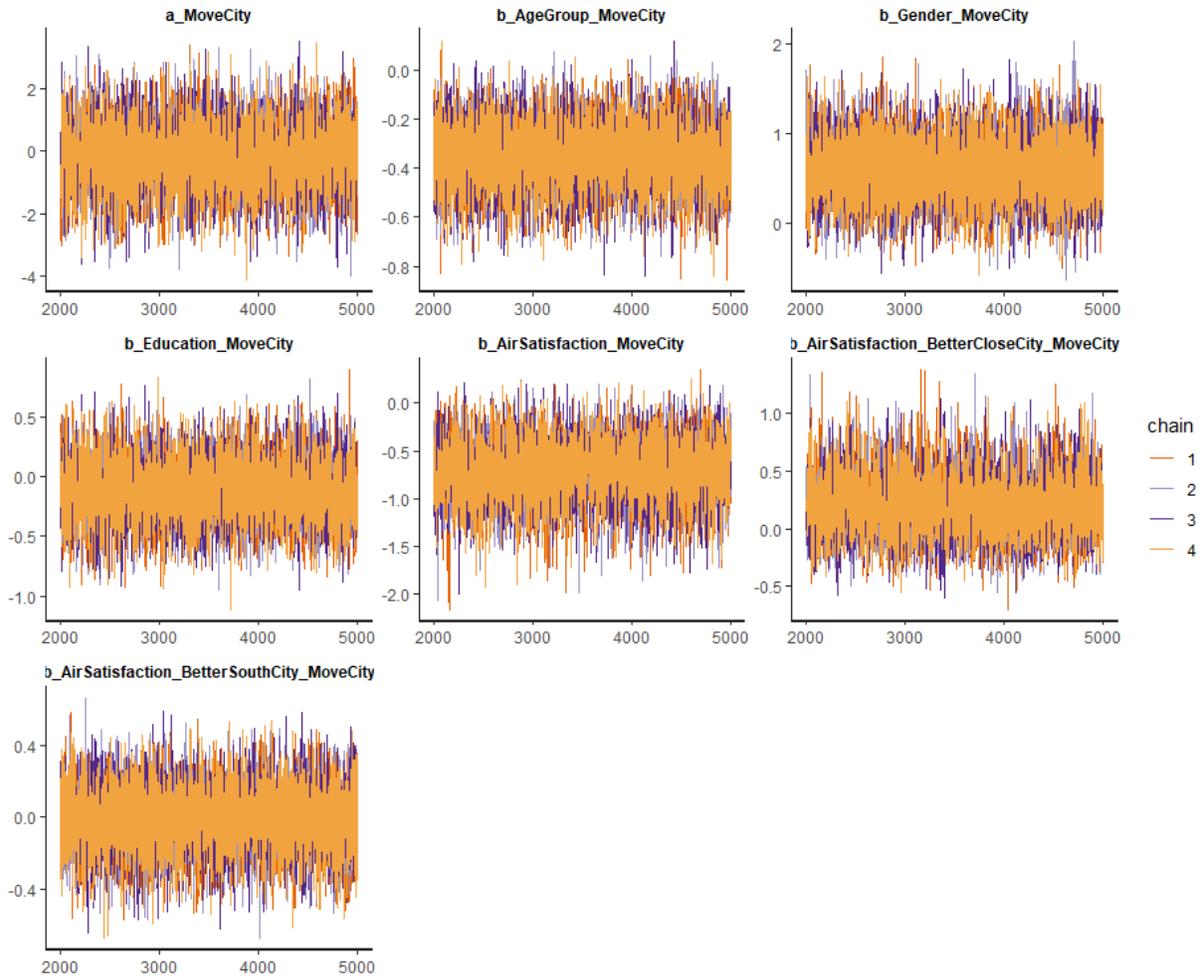

Figure A4: Trace plots for Model 3's posterior parameters

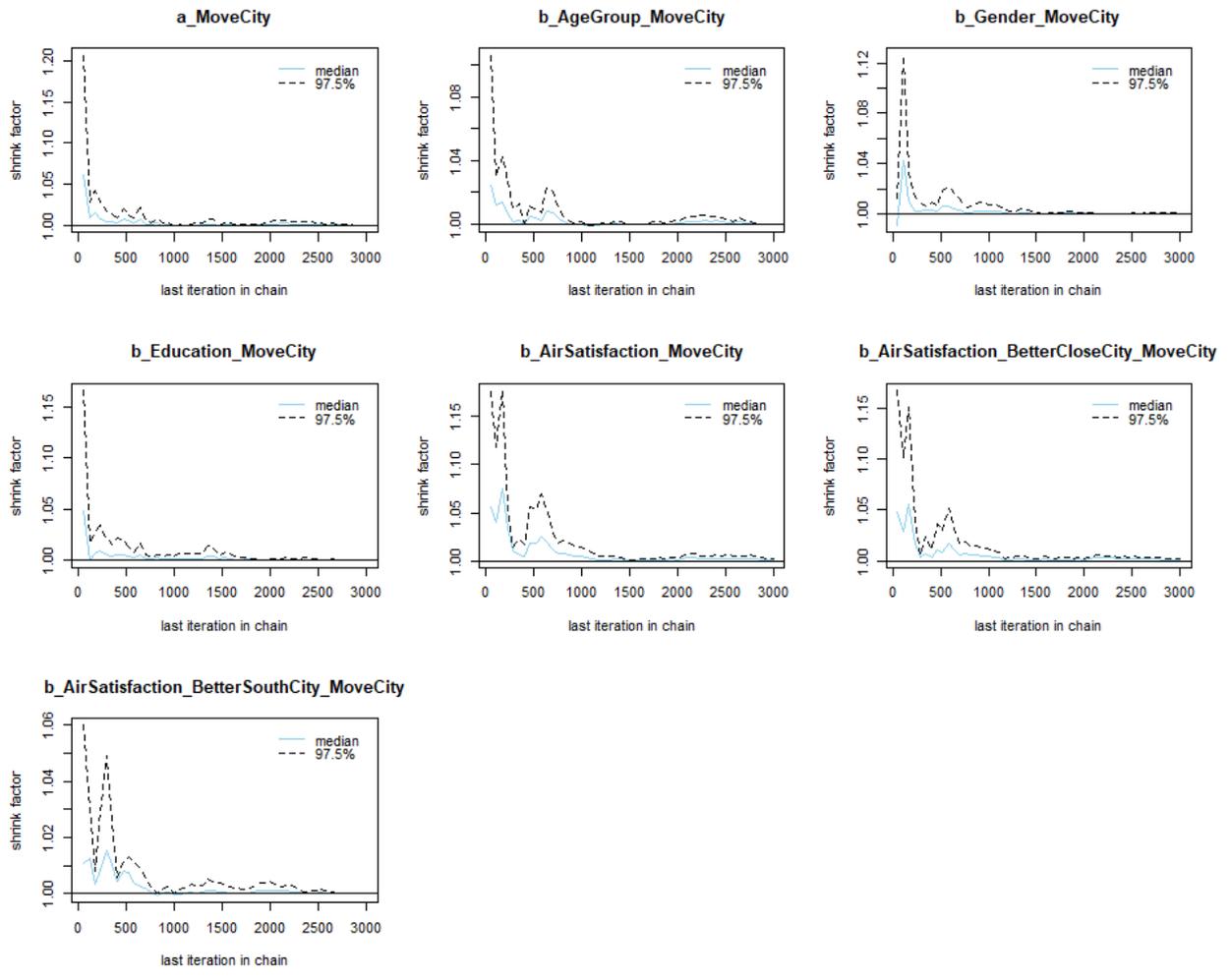

Figure A5: Gelman plots for Model 3's posterior parameters

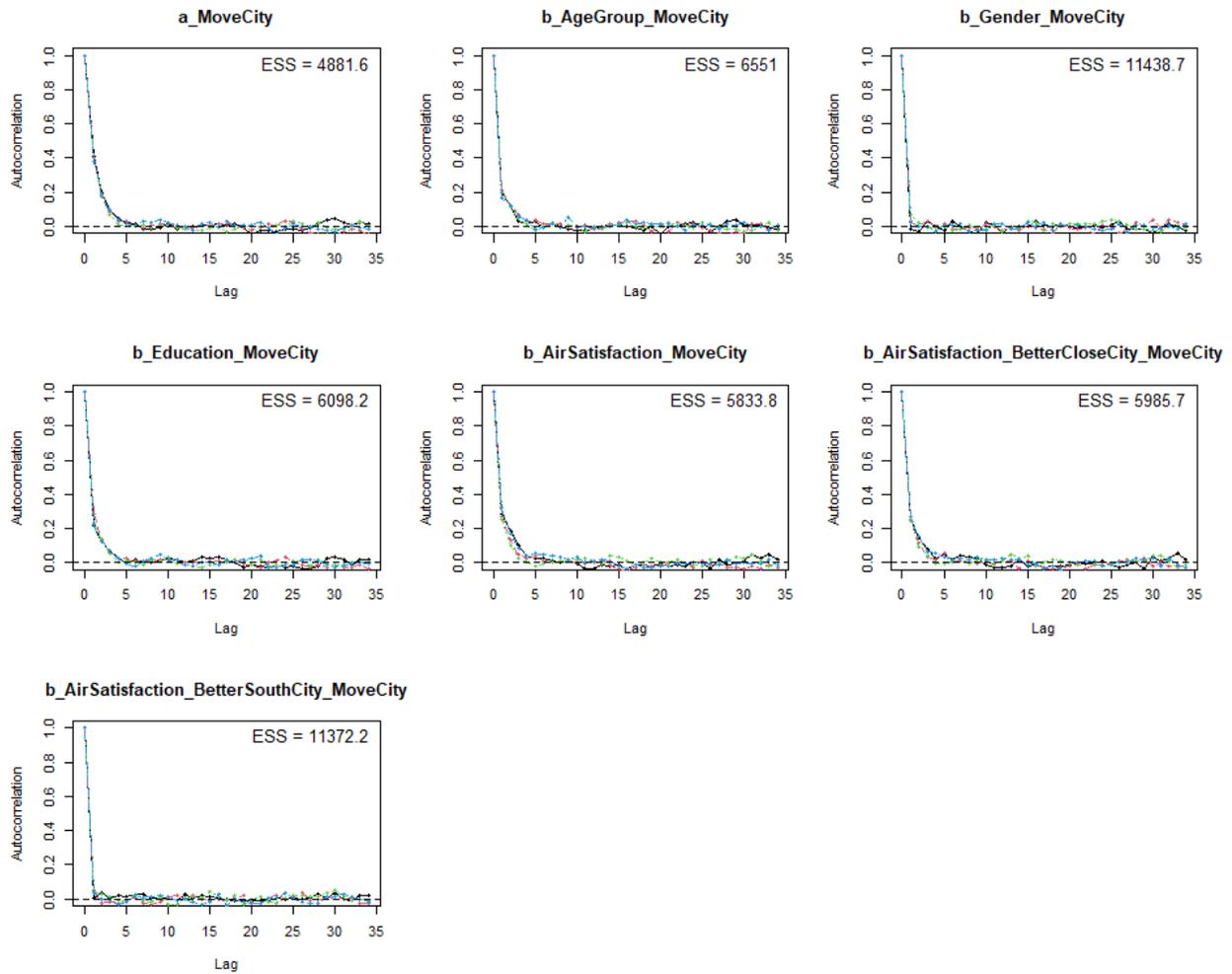

Figure A6: Autocorrelation plots for Model 3's posterior parameters